\documentclass[english,twocolumn,nofootinbib]{revtex4}

\pdfoutput=1

\usepackage[T1]{fontenc}
\usepackage[latin9]{inputenc}
\usepackage{amsmath}
\usepackage{amssymb}
\usepackage{graphicx}
\usepackage{esint}

\makeatletter
\@ifundefined{textcolor}{}
{%
 \definecolor{BLACK}{gray}{0}
 \definecolor{WHITE}{gray}{1}
 \definecolor{RED}{rgb}{1,0,0}
 \definecolor{GREEN}{rgb}{0,1,0}
 \definecolor{BLUE}{rgb}{0,0,1}
 \definecolor{CYAN}{cmyk}{1,0,0,0}
 \definecolor{MAGENTA}{cmyk}{0,1,0,0}
 \definecolor{YELLOW}{cmyk}{0,0,1,0}
}

\usepackage[hidelinks,pdftex]{hyperref}

\makeatother

\usepackage{babel}
\begin{document}

\title{The Resurgence of the Cusp Anomalous Dimension}

\author{In\^{e}s Aniceto}

\email{ines@th.if.uj.edu.pl}

\selectlanguage{english}%

\affiliation{Institute of Physics, Jagiellonian University, \\
Ul. \L{}ojasiewicza 11, 30-348 Krak\'{o}w, Poland.}
\begin{abstract}
This work addresses the resurgent properties of the cusp anomalous
dimension's strong coupling expansion, obtained from the integral
Beisert-Eden-Staudacher (BES) equation. This expansion is factorially
divergent, and its first non-perturbative corrections are related to
the mass gap of the $O(6)$ $\sigma$-model. The factorial divergence
can also be analysed from a resurgence perspective. Building on the
work of Basso and Korchemsky, a transseries ansatz for the cusp anomalous
dimension is proposed and the corresponding expected large-order behaviour
studied. One finds non-perturbative phenomena in both the positive
and negative real coupling directions, which need to be included to
address the analyticity conditions coming from the BES equation. After
checking the resurgence structure of the proposed transseries, it
is shown that it naturally leads to an unambiguous resummation procedure,
furthermore allowing for a strong/weak coupling interpolation.
\end{abstract}
\maketitle

\section{Introduction and Set-up}

The cusp anomalous dimension plays a central role in the study on
many observables in four dimensional gauge theories. In supersymmetric
$\mathcal{N}=4$ Yang--Mills theory (SYM), it appears when studying
the scaling behaviour of the anomalous dimension of a Wilson loop
with a light-like cusp in the integration contour, in the $SL(2)$
sector of the theory \cite{Polyakov:1980ca}. The Wilson loop operators
carry a Lorentz spin $S$ and a twist $L$, and for the case of large
spin and $L\sim\ln S$ the scaling behaviour of the minimal anomalous
dimension is \cite{Belitsky:2006en,Alday:2007mf,Freyhult:2007pz,Frolov:2006qe}
\[
\gamma_{S,L}\left(g\right)=\left(2\Gamma_{\mathrm{cusp}}\left(g\right)+\epsilon\left(g,j\right)\right)\ln S+O\left(L^{-1}\right),
\]
where $(4\pi g)^{2}=\lambda=g_{\mathrm{YM}}^{2}N$ is the 't Hooft
coupling. Also, $j=L/\ln S$ is the only dependence on the twist from
the leading contribution to the scaling, and taking $j=0$ leaves
us with the twist independent cusp anomalous dimension \cite{Korchemsky:1987wg,Korchemsky:1988si}.
It is a function solely of the coupling, and has been thoroughly studied
in different regimes. 

At weak coupling this function can be expanded in powers of $g^{2}$,
with coefficients determined from perturbation theory \cite{Bern:2006ew,Cachazo:2006az},
and the corresponding series is convergent. At strong coupling, through
the AdS/CFT correspondence \cite{Maldacena:1997re}, one can obtain
an expansion in $g^{-1}$ from the semiclassical analysis of the energy
of folded spinning strings in $\mathrm{AdS}{}_{5}\times S^{5}$, where
the Lorentz spin and twist become angular momenta of the string solution
\cite{Gubser:1998bc,Frolov:2002av}. 

Studying the interpolating region between weak and strong coupling
is difficult, and integrability played a crucial role. The all-loop
Bethe ansatz for $\mathcal{N}=4$ SYM \cite{Arutyunov:2004vx,Staudacher:2004tk,Beisert:2005fw}
led to a set of integral equations, the BES equations \cite{Freyhult:2007pz,Eden:2006rx,Beisert:2006ez,Belitsky:2006wg},
describing the anomalous dimension, valid for any arbitrary coupling
(with the FRS equations \cite{Freyhult:2007pz} valid for any scaling
parameter $j$). In terms of an auxiliary function $\gamma\left(2gt\right)$,
the BES equation can be written as
\[
\frac{\gamma(2gt)}{2gt}=K\left(2gt,0\right)-2g\int_{0}^{\infty}\frac{dt'}{e^{t'}-1}K(2gt,2gt')\gamma(2gt').
\]
where the BES kernel $K\left(t,t'\right)$ can be found in \cite{Eden:2006rx,Beisert:2006ez,Basso:2008tx}.
This auxiliary function is related to the cusp anomalous dimension
by
\[
\Gamma_{\mathrm{cusp}}\left(g\right)=8\lim_{t\rightarrow0}\frac{\gamma(2gt)}{2gt}.
\]

Solving these equations at weak coupling returned higher terms of
the convergent perturbative expansion for the small $g\ll1$ region
\cite{Beisert:2006ez}. For intermediate coupling $g\sim1$ a smooth
solution to the BES equation was found numerically \cite{Benna:2006nd}.
At strong coupling different attempts were made at solving the BES
equations \cite{Alday:2007qf,Kostov:2007kx,Beccaria:2007tk}, and
in \cite{Basso:2007wd,Kostov:2008ax} a solution was found leading
to a strong coupling expansion. This approach consisted in noticing
that a change of variables
\[
\Omega\left(t\right)=\gamma(t)\left(1+\cosh\left(\frac{t}{4g}\right)\right),
\]
returns a simpler set of coupled integral equations for $\Omega\left(t\right)$,
which are then solved using Fourier methods \cite{Hasenfratz:1990zz,Hasenfratz:1990ab}.
One subsequently obtains a solution of the BES equation in the form
\cite{Basso:2008tx,Basso:2009gh} 
\[
\Omega\left(\mathrm{i}t\right)=f_{0}\left(t\right)V_{0}\left(t\right)+f_{1}\left(t\right)V_{1}\left(t\right),
\]
where ($x=8\pi g$)
\begin{eqnarray}
f{}_{0}\left(t\right) & \negthickspace\equiv & -1+\\
 &  & \hspace{-7pt}+\sum_{n\ge1}2t\left[c_{+}(n,x)\frac{U_{1}^{+}\left(\frac{nx}{2}\right)}{nx-2t}+c_{-}(n,x)\frac{U_{1}^{-}\left(\frac{nx}{2}\right)}{nx+2t}\right],\nonumber \\
f_{1}\left(t\right) & \negthickspace\equiv & \sum_{n\ge1}nx\left[c_{+}(n,x)\frac{U_{0}^{+}\left(\frac{nx}{2}\right)}{nx-2t}+c_{-}(n,x)\frac{U_{0}^{-}\left(\frac{nx}{2}\right)}{nx+2t}\right].\nonumber 
\end{eqnarray}
The functions $U_{0,1}^{\pm}$ and $V_{0,1}$ can be written in terms
of Whittaker functions of $1^{\mathrm{st}}$ and $2^{\mathrm{nd}}$
kinds, but for our purposes we only need their asymptotic expansions
for large $x$, which can be found in Appendix \ref{sec:App-Asymptotic-Expansions}.
The coefficients $c_{\pm}(n,x)$ are determined from analyticity conditions
on the solution $\Omega\left(\mathrm{i}t\right)$ (given that $\gamma\left(\mathrm{i}t\right)$
is an entire function): from the expresions for $f_{n}(t)$ it already
has the correct pole structure, but one still needs to impose the
existence of zeroes at
\begin{equation}
t_{\textrm{zeroes}}=\frac{x}{2}\alpha_{\ell}\equiv\frac{x}{2}\left(\ell-\frac{1}{4}\right),\quad\ell\in\mathbb{Z}.
\end{equation}

This condition can be re-written as
\begin{eqnarray}
1 & = & \sum_{n\ge1}\frac{c_{+}\left(n,x\right)}{n-\alpha_{\ell}}\left(U_{1}^{+}\left(\frac{nx}{2}\right)\alpha_{\ell}+U_{0}^{+}\left(\frac{nx}{2}\right)n\, r\left(\alpha_{\ell}\right)\right)+\nonumber \\
 &  & \hspace{-15pt}+\sum_{n\ge1}\frac{c_{-}\left(n,x\right)}{n+\alpha_{\ell}}\left(U_{1}^{-}\left(\frac{nx}{2}\right)\alpha_{\ell}+U_{0}^{-}\left(\frac{nx}{2}\right)n\, r\left(\alpha_{\ell}\right)\right),\label{eq:Analyticity-conditions-general}
\end{eqnarray}
where $r\left(\alpha\right)$ is the ratio of functions $V_{1}\left(nx/2\right)$
and $V_{0}\left(nx/2\right)$, as defined in Appendix \ref{sec:App-Asymptotic-Expansions}.
This analyticity condition allows us to determine the coefficients
$c_{\pm}\left(n,x\right)$ order by order as expansions in large coupling
$x$. Once these coefficients are known, the cusp anomalous dimension
is given by ($x=8\pi g$) 
\begin{eqnarray}
\frac{\Gamma(g)}{2g} & = & 1-2f_{1}\left(0\right)\label{eq:Cusp-definition}\\
 &  & \hspace{-25pt}=1-2\sum_{n\ge1}\left[c_{+}(n,x)\, U_{0}^{+}\left(\frac{nx}{2}\right)+c_{-}(n,x)\, U_{0}^{-}\left(\frac{nx}{2}\right)\right].\nonumber 
\end{eqnarray}

The strong coupling expansion found in this way is asymptotic. Moreover,
the series is non-Borel summable for positive real coupling, due to
the existence of singularities on the positive real axis of the Borel
plane, which give rise to non-perturbative, exponentially suppressed
corrections at strong coupling. In order to understand the analytic
properties of the solution to the BES equation at strong coupling,
one needs to account for all the non-perturbative phenomena in this
limit. In \cite{Basso:2008tx,Basso:2009gh} the above procedure was
taken a step further and the perturbative coefficients around the
first non-perturbative correction were determined. 

Both scaling function $\epsilon\left(g,j\right)$ and cusp anomalous
dimension $\Gamma_{\mathrm{cusp}}\left(g\right)$ have non-perturbative
corrections. In \cite{Alday:2007mf} it was proposed that the scaling
function $\epsilon\left(g,j\right)$ at strong coupling is directly
related to the energy density of the ground state of the $O(6)$ non-linear
$\sigma$-model embedded in $\mathrm{AdS}{}_{5}\times S^{5}$ (taking
$j/2$ to be the particle density). Moreover, the non-perturbative
corrections appearing in $\epsilon\left(g,j\right)$ at strong coupling
are given by the mass scale (mass gap) of the $O(6)$ model. Agreement
between these two quantities was checked in \cite{Basso:2008tx,Bajnok:2008it,Volin:2009wr},
at the level of the first non-perturbative correction to the scaling
function. As for the cusp anomalous dimension, as it solves a different
integral equation altogether, such a relation was less expected. Nevertheless,
in \cite{Basso:2009gh}, it was shown that the first non-perturbative
correction to the anomalous dimension is exactly given by the square
of the $O(6)$ mass gap.

Two important questions still remain at this point: are we aware of
all of the non-perturbative phenomena defining the analytic properties
of the cusp anomalous dimension? How can we systematically deal with
a non-Borel summable asymptotic series? To answer both these questions
we will now turn to the theory of resurgence.

Resurgent functions have been seen in a wide range of systems. In
mathematics they appear for example as solutions of differential and
finite difference equations (see \textit{e.g.} the well studied cases
of Painlev\'{e} I, II and Riccati non-linear differential equations
\cite{Daalhuis05,Garoufalidis:2010ya,Aniceto:2011nu,Schiappa:2013opa}).
Analogously, often one can only determine physical observables in
specific regimes of the coupling of the theory via a series expansion
such as
\begin{equation}
\left\langle \mathcal{O}\left(g\right)\right\rangle \simeq\sum_{k\ge0}\mathcal{O}_{k}g^{-k}.
\end{equation}
However, these expansions are often asymptotic: the coefficients are
factorially divergent, with large order behaviour
\begin{equation}
\mathcal{O}_{k}\sim\frac{\Gamma\left(k+\beta\right)}{A^{k+\beta}}\,,\quad k\gg1.
\end{equation}
$A,\beta$ are numbers related to the position and type of singularities
of the related Borel transform.

It is well known that this divergence hints to the existence of non-perturbative
phenomena unaccounted for in the perturbative series expansions. In
physical settings, the existence of non-perturbative phenomena has
been long noticed in the contexts of quantum mechanics \cite{Bender:1969si,Bender:1990pd}
and quantum field theories \cite{Dyson:1952tj}, associated to instantons
\cite{ZinnJustin:1980uk} and renormalons \cite{Beneke:1998ui}. In
these examples, the existence of asymptotic multi-instanton sectors
allowed for a complete unambiguous description of the energy eigenvalues
via a transseries solution and resurgence \cite{Bogomolny:1980ur,ZinnJustin:1981dx,ZinnJustin:1982td}.
Since then, the asymptotic behaviour of perturbation theory and the
resurgence behind it was seen to exist in many different examples
in physical systems, from quantum mechanics \cite{Dunne:2013ada,Basar:2013eka,Dunne:2014bca},
to large $N$ gauge theories \cite{David:1990sk,David:1992za,Marino:2006hs,Marino:2007te,Marino:2008ya,Marino:2008vx,Pasquetti:2009jg,Aniceto:2011nu,Schiappa:2013opa,Couso-Santamaria:2015wga},
quantum field theories \cite{Dunne:2012ae,Dunne:2012zk,Cherman:2013yfa,Dunne:2013ada,Aniceto:2014hoa,Shifman:2014fra,Basar:2015xna,Dunne:2015ywa}
and topological strings \cite{Santamaria:2013rua,Couso-Santamaria:2014iia,Grassi:2014cla}. 

To account for all non-perturbative phenomena, one upgrades our perturbative
expansions into a transseries \cite{Marino:2008ya}: a formal expansion
in both perturbative variable $g$ and non-perturbative monomials $\mathrm{e}^{-Ag}$.
Schematically 
\begin{equation}
\left\langle \mathcal{O}\left(g,\sigma\right)\right\rangle =\sum_{n\ge0}\sigma^{n}\mathrm{e}^{-nAg}g^{\beta_{n}}\sum_{k\ge0}\mathcal{O}_{k}^{(n)}g^{-k},\label{eq:intro-transseries}
\end{equation}
where $\sigma$ is a parameter to be fixed from some boundary conditions
specific to each problem. The transseries is a formal object, since
for each non-perturbative sector labeled by $n$ one has an associated
asymptotic expansion $\Phi_{n}\left(g\right)\simeq g^{\beta_{n}}\sum_{k\ge0}\mathcal{O}_{k}^{(n)}g^{-k}$,
with coefficients growing factorially at large orders. However, these
sectors are not independent of each other: they are sectors of a resurgent
transseries, whose large-order growth is intimately related \cite{Aniceto:2011nu}.
A resurgent transseries is an expansion like (\ref{eq:intro-transseries}),
where the coefficients of one sector $\mathcal{O}_{k}^{(m)}$ are
related to, \textit{i.e. resurge} in, the coefficients of neighbouring
sectors $\mathcal{O}_{k}^{(m')}$ ($m$ close to $m'$). For example,
for the perturbative sector ($n=0$), one expects a direct relation
to the $n=1$ sector
\begin{equation}
\mathcal{O}_{k}^{(0)}\sim\frac{\Gamma\left(k+\beta_{n}\right)}{A^{k+\beta_{n}}}\left(\mathcal{O}_{0}^{(1)}+\frac{A}{g}\mathcal{O}_{1}^{(1)}+\cdots\right)\,,\; k\gg1.
\end{equation}
The exact expressions for these large-order relations \cite{Aniceto:2011nu}
can be determined via resurgent analysis (for a introduction to resurgence
see \cite{Aniceto:2011nu,sauzin14,Dorigoni:2014hea,Upcoming:2015}
and references therein). The associated Borel transforms $\mathcal{B}\left[\Phi_{n}\right](s)$,%
\footnote{Borel transforms are determined by inverse Laplace transforms to each
term in the expansion, or equivalently $g^{-k}\rightarrow s^{k-1}/\Gamma(k)$.%
} have a non-zero radius of convergence and singularities on the corresponding
Borel $s$-plane at positions $s=nA,\, n\in\mathbb{N}$. 

At this point we have a formal solution for our observable, and we
now need to retrieve physical information from the asymptotic series
$\Phi_{n}(g)$. This is done via Borel resummation: the calculated
Borel transform has a non-zero radius of convergence, and one can
determine the analytic function associated with each series $\mathcal{B}\left[\Phi_{n}\right](s)$,
either exactly or by finding an approximate analytic result via the
so-called Borel--Pad\'{e} approximants \cite{Bender:1978,Marino:2008ya,Aniceto:2011nu}.
Once the function or its approximant is known we then perform a resummation
via a Lapace transform
\begin{equation}
\mathcal{S}\Phi_{n}\left(g\right)=\int_{0}^{+\infty}ds\,\mathrm{e}^{-s\, g}\mathcal{B}\left[\Phi_{n}\right](s),
\end{equation}
and the full answer for the observable is given by the transseries
with each of its sectors resummed. This can only be performed if no
poles exist in the direction of integration on the Borel plane, in
this case on the positive real line. If instead $A$ is positive and
real, the positive real line is called a Stokes line (singular direction
on the Borel plane), and the series is said to be non-Borel summable:
only lateral resummations can be defined: 
\begin{equation}
\mathcal{S}_{\pm}\Phi_{n}\left(g\right)=\int_{0}^{+\infty\,\mathrm{e}^{\pm\mathrm{i}\epsilon}}ds\,\mathrm{e}^{-s\, g}\mathcal{B}\left[\Phi_{n}\right](s),\label{eq:Lateral-resums}
\end{equation}
these lateral resummations differ by a non-perturbative ambiguity
$\left(\mathcal{S}_{+}-\mathcal{S}_{-}\right)\Phi_{n}\left(g\right)\sim\mathrm{e}^{-Ag}$,
which is purely imaginary when the coefficients $\mathcal{O}_{k}^{(n)}$
are real and the Stokes line is along the real axis. Now the importance
of having a resurgent transseries becomes apparent: due to the relations
between different sectors, by taking into account the full transseries
and a specific value for the transseries parameter $\sigma$, the
ambiguities between different sectors cancel each other, and one
is left with a non-ambiguous real-valued result. This is called median
resummation (see \cite{Aniceto:2013fka} and references therein). 

Recalling the transseries (\ref{eq:intro-transseries}), let us assume
that the positive real axis is a Stokes line and choose the lateral
Borel resummation $\mathcal{S}_{+}$ for every sector of the transseries.
This resummation will have real and imaginary parts
\begin{eqnarray}
\mathcal{S}_{+}\Phi_{n} & = & \frac{1}{2}\left(\mathcal{S}_{+}+\mathcal{S}_{-}\right)\Phi_{n}+\frac{1}{2}\left(\mathcal{S}_{+}-\mathcal{S}_{-}\right)\Phi_{n}\nonumber \\
 & \equiv & \mathcal{S}_{R}\Phi_{n}+\mathrm{i}\mathcal{S}_{I}\Phi_{n},
\end{eqnarray}
The imaginary contribution $\mathcal{S}_{I}\Phi_{n}$ is just the
ambiguity coming from the sector $\Phi_{n}$. We can now determine
the real and imaginary parts of the resummed transseries \cite{Aniceto:2013fka}
\begin{eqnarray*}
\mathcal{S}_{+}\left\langle \mathcal{O}\left(g,\sigma\right)\right\rangle  & \equiv & \sum_{n\ge0}\sigma^{n}\mathrm{e}^{-nAg}\mathcal{S}_{+}\Phi_{n}(g)\equiv\sum_{n\ge0}\sigma^{n}F^{(n)}\left(g\right)\\
 & \equiv & \mathcal{S}_{R}\left\langle \mathcal{O}\right\rangle +\mathrm{i}\,\mathcal{S}_{I}\left\langle \mathcal{O}\right\rangle .
\end{eqnarray*}
 The ambiguity in the resummation of the transseries is just its imaginary
part ($\sigma=\sigma_{R}+\mathrm{i}\,\sigma_{I}$) 
\begin{eqnarray}
\mathcal{S}_{I}\left\langle \mathcal{O}\right\rangle  & = & \mathrm{Im}\left(F^{(0)}\right)+\sigma_{I}\mathrm{Re}\left(F^{(1)}\right)+\sigma_{R}\mathrm{Im}\left(F^{(1)}\right)+\nonumber \\
 &  & \hspace{-45pt}+2\sigma_{R}\sigma_{I}\mathrm{Re}\left(F^{(2)}\right)+\left(\sigma_{R}^{2}-\sigma_{I}^{2}\right)\mathrm{Im}\left(F^{(2)}\right)+\cdots.\label{eq:Imaginary-part-transseries}
\end{eqnarray}
Median resummation is a specific prescription to cancel this imaginary
contribution to the resummation of the transseries along the positive
real axis to all orders, by some carefully chosen values of the transseries
parameter $\sigma=\sigma_{0}$.%
\footnote{In simple cases it was seen that $\mathrm{i}\sigma_{I}=\frac{S_{1}}{2}$
was enough to cancel the ambiguity \cite{Aniceto:2013fka}, with residual
freedom left in the real part $\sigma_{R}$.%
} This cancelation happens to all orders, and we are left with a real
unambiguous answer 
\begin{eqnarray}
\mathcal{S}_{R}\left\langle \mathcal{O}(g,\sigma_{0})\right\rangle  & = & \mathrm{Re}\left(F^{(0)}\right)+\sigma_{0,R}\mathrm{Re}\left(F^{1}\right)+\label{eq:Real-part-transseries}\\
 &  & +\left(\sigma_{0,R}^{2}-\sigma_{0,I}^{2}\right)\mathrm{Im}\left(F^{(2)}\right)+\cdots.\nonumber 
\end{eqnarray}
This resummed result can then be interpolated from the original regime
where the asymptotic series were defined, into any complex value of
the coupling $g$, taking into consideration any crossing of singular
Stokes lines. The systematic resummation and interpolation from asymptotic
series using resurgence has recently been addressed for different
problems \cite{Grassi:2014cla,Couso-Santamaria:2015wga,Basar:2015xna,Heller:2015dha,Hatsuda:2015owa}.

The aim of this paper is to perform a resurgent analysis of the expansions
found in \cite{Basso:2009gh} for the strong coupling regime of the
cusp anomalous dimension. We start from the solution to the BES equations
presented above, and enforce the analytic properties at the level
of the expansion coefficients. We then determine the structure of
singularities of the Borel transform associated to the perturbative
sector by means of a Borel--Pad\'{e} approximant. This allows us
to finally propose a transseries ansatz for the cusp anomalous dimension
which encompasses all the expected non-perturbative phenomena existing
at strong coupling. Using this ansatz in the analyticity conditions,
we determine the coefficients of our transseries, solved order by
order for every sector.

Equipped with the series expansion for perturbative and non-perturbative
sectors of the cusp anomalous dimension, we then check its resurgent
properties via the large-order relations. Along the way we determine
the relevant Stokes constant associated with the Stokes transition
across the positive real line.

We finish by using the methods of median resummation to systematically
define a non-ambiguous resummed result valid at any value of the coupling,
which encodes the analytic properties of the solution and can be used
to interpolate between strong and weak coupling regimes.

\section{Singularity Structure of the Cusp}

In the interest of finding the correct transseries solution for the
cusp anomalous dimension, we first analyse its perturbative asymptotic
series. With that goal in mind we assume the coefficients $c_{\pm}\left(n,x\right)$
have a simple (asymptotic) expansion in powers of $x^{-1}$ where
$x=8\pi g$:
\begin{equation}
c_{\pm}\left(n,x\right)=x^{\beta_{\pm}\left(0\right)}n^{\pm1/4}\sum_{k=0}^{+\infty}\phi_{k}^{(0,\pm)}\left(n\right)\, x^{-k}.\label{eq:coeffs-cpm-pert}
\end{equation}
Substituting this ansatz into the analyticity conditions (\ref{eq:Analyticity-conditions-general}),
and making use of the asymptotic expansions in Appendix \ref{sec:App-Asymptotic-Expansions},
as well as properties of sums found in Appendix \ref{sec:App-Relations-Between-Sums},
one obtains for each power in $x^{-k}$ relations for the coefficients
$\phi_{k}^{(0,\pm)}(n)$ depending on the ones for lower $k$. Solving
these relations iteratively (in the same way as was done in \cite{Basso:2009gh}
for the first few coefficients), we determine 
\begin{equation}
\phi_{k}^{(0,\pm)}(n)=\sum_{m=0}^{n}Q_{k,m}^{(0,\pm)}n^{-n}\phi_{0}^{(0,\pm)}(n),
\end{equation}
where the expressions for $\phi_{0}^{(0,\pm)}(n)$ are very simple
and can be found in Appendix \ref{sec:App-Relations-Between-Sums}.
The analyticity condition imposes restrictions on the coefficients
$\beta_{\pm}(0)$:
\begin{equation}
\beta_{\pm}(0)=\pm1/4.
\end{equation}
The general solution for the numerical coefficients $Q_{k,m}^{(0,\pm)}$
is then simply given by 
\begin{eqnarray}
4Q_{k,k}^{(0,+)} & = & -\sum_{r=0}^{k-1}Q_{r,r}^{(0,+)}\sum_{\ell=0}^{k-r}M_{k-r,\ell}^{(0,+)},\\
4Q_{k,k}^{(0,-)} & = & -\sum_{r=0}^{k-1}Q_{r,r}^{(0,-)}\sum_{\ell=0}^{k-r}M_{k-r,\ell}^{(1,-)},\nonumber 
\end{eqnarray}
for $m=k$, and for $0\le a<k$
\begin{eqnarray}
2Q_{k,a}^{(0,+)} & = & -4\sum_{m=a+1}^{k}Q_{k,m}^{(0,+)}\, K_{m-a}^{(0,+)}-\\
 &  & -\sum_{r=1}^{a}\sum_{\ell=0}^{r}M_{r,r-\ell}^{(0,+)}\sum_{m=a}^{k}Q_{k-r,m-r}^{(0,+)}K_{m-a}^{(0,+)}-\nonumber \\
 &  & -\sum_{r=a+1}^{k}\sum_{m=r}^{k}Q_{k-r,m-r}^{(0,+)}K_{m-a}^{(0,+)}\sum_{\ell=0}^{a}M_{r,r-\ell}^{(0,+)}+\nonumber \\
 &  & +\sum_{r=a+1}^{k}\sum_{m=r}^{k}Q_{k-r,m-r}^{(0,-)}K_{m-a}^{(0,-)}M_{r-1,r-a-1}^{(1,+)},\nonumber 
\end{eqnarray}
where similar solutions can be written for $Q_{k,a}^{(0,-)}$ by exchanging
$K_{m}^{(0,+)}\leftrightarrow K_{m}^{(0,-)}$, and $M_{r,\ell}^{(0,+)}\rightarrow M_{r,\ell}^{(1,-)}$,
$M_{r,\ell}^{(1,+)}\rightarrow M_{r,\ell}^{(0,-)}$. The definitions
for the coefficients $M_{r,\ell}^{(n,\pm)}$ and $K_{m}^{(n,\pm)}$
can be found in the Appendices. One can now determine several coefficients
in the expansions (\ref{eq:coeffs-cpm-pert}), which was done numerically
up to $k=200$. For the cusp anomalous dimension one then uses the
expansion (\ref{eq:Cusp-definition}), and the expansions for the
functions $U_{k}^{\pm}\left(\frac{nx}{2}\right),\, k=0,1$ given in
Appendix \ref{sec:App-Asymptotic-Expansions}. 

The strong coupling perturbative expansion for the cusp anomalous
dimension becomes 
\begin{equation}
\Gamma_{\mathrm{cusp}}(g)=2g\left(1+\Gamma^{(0)}\left(4\pi g\right)+O\left(\mathrm{e}^{-\frac{1}{2}4\pi g}\right)\right),
\end{equation}
where the perturbative asymptotic expansion is
\begin{equation}
\Gamma^{(0)}(x)\equiv-2f_{1,\textrm{pert}}(0)\simeq\sum_{k=1}^{+\infty}\left(\frac{x}{2}\right)^{-k}\Gamma_{k}^{(0)}.
\end{equation}
The expansion coefficients are simply given by
\begin{eqnarray}
\Gamma_{k}^{(0)} & \simeq & 2^{1-k}\sum_{m=0}^{k-1}\sum_{s=0}^{m}S_{+}\left(s\right)Q_{k-1-s,m-s}^{(0,+)}K_{m+1}^{(0,+)}+\label{eq:Cusp-pert-sol}\\
 &  & +2^{1-k}\sum_{m=0}^{k-1}\sum_{s=0}^{m}S_{-}\left(s\right)Q_{k-1-s,m-s}^{(0,-)}K_{m+1}^{(0,-)},\nonumber 
\end{eqnarray}
with 
\begin{eqnarray}
S_{+}\left(s\right) & = & \frac{\Gamma\left(5/4+s\right)\Gamma\left(1/4+s\right)}{\Gamma\left(5/4\right)\Gamma\left(1/4\right)\, s!}(-1)^{s},\\
S_{-}(s) & = & \frac{\Gamma\left(3/4+s\right)\Gamma\left(-1/4+s\right)}{\Gamma\left(3/4\right)\Gamma\left(-1/4\right)\, s!}(-1)^{s}.\nonumber 
\end{eqnarray}

One can now see that the coefficients $\Gamma_{k}^{(0)}$ grow factorially
for large order $k$. In fact they grow as $\Gamma(k-1/2)$, in agreement
with the factorial growth found in \cite{Basso:2007wd}. Associated
with this factorial growth there will be non-perturbative phenomena
dictating the asymptotic nature of the series. This non-perturbative
phenomena is most naturally represented as singularities on the complex
Borel plane: we expect to find singularities at positions $s=n\, A$,
where $A=1/2$ -- these will be associated to non-perturbative exponentially
suppressed contributions of the type $\mathrm{e}^{-n\, A\,4\pi g}$. 

\begin{figure}
\begin{centering}
\includegraphics[width=1\columnwidth]{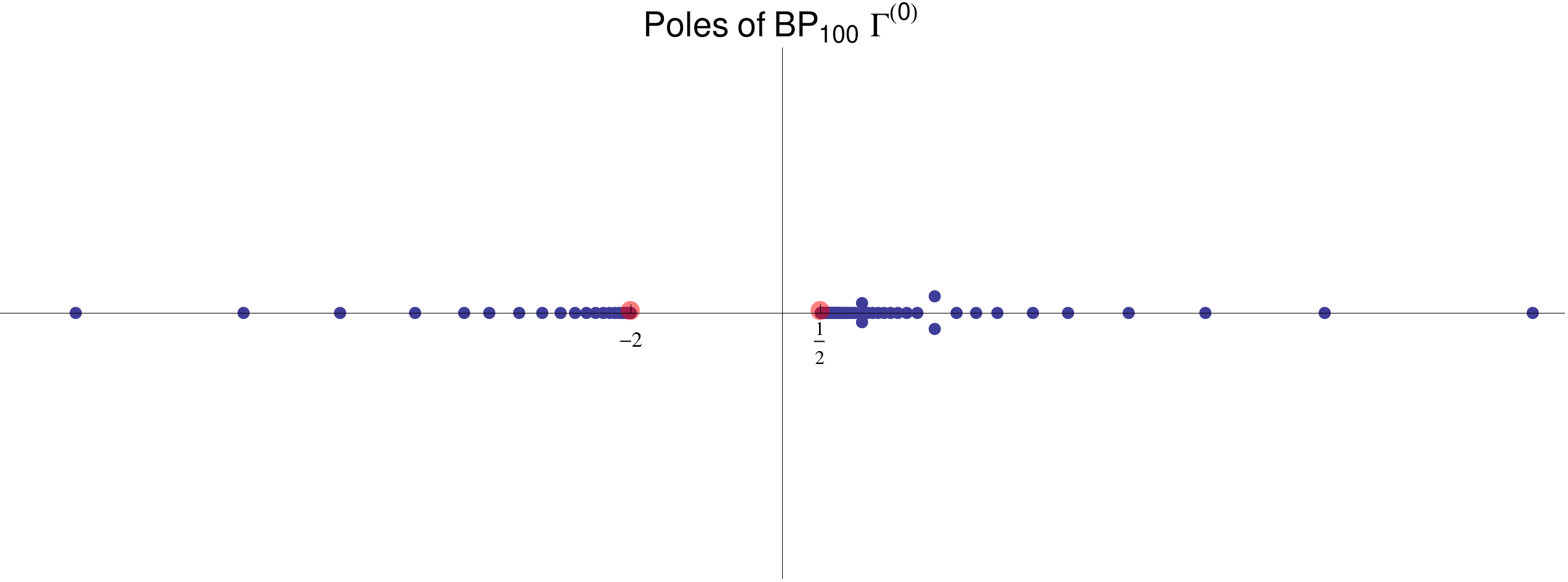}
\par\end{centering}

\protect\caption{Poles of the diagonal Borel--Padé approximant of order 100 for the
perturbative series of $\Gamma(2g)/(2g)-1$. There accumulation of
poles in both positive and negative real directions, starting at $s=A\equiv1/2$
and at $s=-4A=-2$. Note the existence of spurious poles away from
the real line: non-stable numerical effects of the Pad\'{e} method,
which move away by choosing different non-diagonal approximants.\label{fig:Poles-of-Pert-Borel-Pade}}
\end{figure}

To study the singularity structure associated with the perturbative
expansion $\Gamma^{(0)}(g)$ we determine its Borel transform via
the usual approach
\begin{equation}
\mathcal{B}\left[\Gamma^{(0)}\right](s)=\sum_{k=0}^{+\infty}\frac{\Gamma_{k+1}^{(0)}\, s^{k}}{\Gamma\left(k+1\right)}.
\end{equation}
This expansion will now have a non-zero radius of convergence and
we approximate the corresponding function via the method of Pad\'{e}
approximants: using a diagonal approximant or order $N=100$ (half
the order of coefficients calculated for the original series), we
determine the best fit for a ratio of two polynomials $BP_{N}\left(\Gamma^{(0)}\right)$
and analyse the structure of poles for this function. This allows
us to see the position and type of nearest singularities to the origin
on the Borel plane: in particular, a condensation of poles hints to
the existence of a branch cut. In Figure \ref{fig:Poles-of-Pert-Borel-Pade}
the structure of poles of the Borel--Pad\'{e} approximant is given.
We find the expected pole at $s=A=1/2$, but we also find another
type of singularity on the negative real axis at $s=-4A=-2$.

The first type of singularity had already been known, it is directly
related to the square of the mass gap of the $O(6)$ $\sigma$-model
embedded in $\mathrm{AdS}{}_{5}\times S^{5}$ \cite{Basso:2009gh}.
The fact that it lies on the positive real axis prevents us from defining
a resummation on this axis: we can only define lateral resummations
(\ref{eq:Lateral-resums}) which differ by an imaginary ambiguous
contribution. However, the second type of singularity lies on the
negative real axis. Even though it will not give any ambiguous contribution
to a resummation for real and positive coupling $g$, the analyticity
conditions will not be blind to it. These types of singularities have
been found before in the study of the Painlev\'{e} I and II equations
\cite{Garoufalidis:2010ya,Aniceto:2011nu,Schiappa:2013opa}.

If the perturbative expansion for some observable is asymptotic, one
should upgrade the solution to include non-perturbative sectors, into
what is called a transseries. The most important ingredient to writing
a transseries solution fully describing our observable is to include
all possible sectors associated with singularities on the Borel plane.
In the present case, this means upgrading the expansions for $c_{\pm}(n,g)$
to a transseries including both types of singularities found. 

\section{Transseries and Analyticity Conditions\label{sec:Transseries-and-Analyticity}}

We can now write a trasseries ansatz for our coefficients $c_{\pm}\left(n,g\right)$
appearing in (\ref{eq:Cusp-definition}). 
\begin{equation}
c_{\pm}\left(n,x;\sigma_{i}\right)=\hspace{-5pt}\sum_{m_{1},m_{2}=0}^{+\infty}\sigma_{1}^{m_{1}}\sigma_{2}^{m_{2}}\mathrm{e}^{(4m_{2}-m_{1})A\frac{x}{2}}c_{\pm}^{(m_{i})}\left(n,x\right),\label{eq:Transseries-cpm}
\end{equation}
where $\sigma_{i}$ are the transseries (or instanton counting) parameters,
$A=\frac{1}{2}$ and $c_{\pm}^{(m)}(n,x)$ are perturbative expansions
around each non-perturbative sector $(m_{1},m_{2})$: $c_{\pm}^{(0)}(n,x)$
are simply the perturbative expansions (\ref{eq:coeffs-cpm-pert})
found in the previous Section, while the others will generically be
\begin{equation}
c_{\pm}^{(m_{i})}\left(n,x\right)\simeq x^{\beta_{\pm}\left(m_{i}\right)}\sum_{k=0}^{+\infty}x^{-k}\phi_{k}^{(m,\pm)}\left(n\right)n^{\pm1/4},\label{eq:transseries-pert-sectors-cpm}
\end{equation}
where $\beta_{\pm}\left(m\right)$ are numerical factors associated
with the type of branch cuts on the Borel plane. In general one is
interested in the transseries solution at some particular value of
the coupling $x$ (usually real positive), and the parameters $\sigma_{i}$
are fixed by some physical input. Moreover, having a non-zero $\sigma_{2}$
for positive real $x$ would lead to unstable, exponentially enhanced
contributions, which physically we know should not be included. Nevertheless,
such sectors need to be introduced in order to account for all the
analytic properties of the problem at hand, even if in the end, once
resummation is performed, the parameter $\sigma_{2}$ gets fixed to
zero.

Another important issue is that of resonance \cite{Garoufalidis:2010ya,Aniceto:2011nu,Schiappa:2013opa,Upcoming:2015}:
for transseries with more than one type of exponential behaviour $\mathrm{e}^{-m_{1}A_{1}-m_{2}A_{2}}$,
if there are values $m_{i}$ such that $m_{1}A_{1}+m_{2}A_{2}=0$,
a phenomena called resonance occurs. In the case of transseries solutions
to non-linear differential equations, it is seen that the associated
recursion relations break down at these locations unless we enhance
the perturbative expansions $c_{\pm}^{(m)}$ (\ref{eq:transseries-pert-sectors-cpm})
to include other non-perturbative sectors, such as some finite number
of powers of $\log(x)$. In the present case we will find resonance
when $m_{1}=4m_{2}$, and one can expect a rich structure like the
one found for the Painlev\'{e} solutions \cite{Aniceto:2011nu,Schiappa:2013opa}.
For our present case we will limit the study to the transseries contributions
up to \textquotedbl{}two instantons\textquotedbl{}, $m=2$ in (\ref{eq:Transseries-cpm}),
and will not reach these structures in our analysis.

Having written the transseries ansatz, we now need to substitute it
in the analyticity conditions (\ref{eq:Analyticity-conditions-general}).
After some algebra, we can re-write these conditions as 
\begin{eqnarray}
2 & = & \sum_{m_{2},m=0}^{+\infty}\sigma_{1}^{m}\sigma_{2}^{m_{2}}\mathrm{e}^{-A\frac{x}{2}\left(m-4m_{2}\right)}\times\label{eq:Analyticity-cond-transseries}\\
 &  & \times\sum_{r=0}^{\left\lfloor \frac{m}{4\left|\alpha\right|}\right\rfloor }\sigma_{1}^{-4\left|\alpha\right|r}\left(F_{+}^{(m,r)}\left(n,x\right)+F_{-}^{(m,r)}\left(n,x\right)\right),\nonumber 
\end{eqnarray}
which need to be obeyed for every zero $\alpha=\ell-\frac{1}{4}$,
$\ell\in\mathbb{Z}$. Throughout this paper, we focus on the contributions
with $m\le2$ and $m_{2}=0$. Therefore we will leave the issue of
resonance and higher non-perturbative corrections for subsequent work.
We briefly note that the last sum in (\ref{eq:Analyticity-cond-transseries})
goes up to the integer part $\left\lfloor \frac{m}{4\left|\alpha\right|}\right\rfloor $,
which will be non-zero if $m\ge4\left|\alpha\right|$. For $\alpha_{\ell}<0$
this already happens for $m=1$ and $\alpha_{0}=-1/4$. For $\alpha_{\ell}\ge\alpha_{1}=3/4>0$,
the first non-perturbative contribution will be at $m=3$. The case
$m=4$ is somewhat special, as both $\alpha_{0}=-1/4$ and $\alpha_{1}=3/4$
return a non-zero sum (for $m\ge5$ there will be two negative zeroes).
It would also be at this point (taking $m_{2}=1$) that we find the
first instance of resonance: it is likely that the two effects will
mix in the analyticity conditions.

Taking $m_{2}=0$, then the functions $F_{\pm}$ are%
\footnote{See Appendices for the expansions used, and $\bar{R}_{k}\left(x,\alpha\right)=R_{k}\left(x,\left|\alpha\right|\right)$
if $\alpha>0$ while $\bar{R}_{k}\left(x,\alpha\right)=\widetilde{R}_{k}\left(x,\left|\alpha\right|\right)$
if $\alpha<0$.%
} 
\begin{eqnarray}
F_{+}^{(m,r)}\left(n,x\right) & = & x^{\beta_{+}\left(m-4\left|\alpha\right|r\right)-1/4}\times\\
 &  & \times\sum_{k=0}^{+\infty}x^{-k}\sum_{n\ge1}\frac{\phi_{k}^{(m-4\left|\alpha\right|r,+)}\left(n\right)}{n-\alpha}\times\nonumber \\
 &  & \hspace{-20pt}\hspace{-20pt}\hspace{-20pt}\times A_{0,1}\left(x,-n\right)\left(\delta_{r,0}\frac{A_{1,1}\left(x,-n\right)}{A_{0,1}\left(x,-n\right)}\alpha+x^{-1}\bar{R}_{r}\left(x,\alpha\right)\right),\nonumber 
\end{eqnarray}
\begin{eqnarray}
F_{-}^{(m,r)}\left(n,x\right) & = & x^{\beta_{-}\left(m-4\left|\alpha\right|r\right)-3/4}\times\\
 &  & \times\sum_{k=0}^{+\infty}x^{-k}\sum_{n\ge1}\frac{\phi_{k}^{(m-4\left|\alpha\right|r,-)}\left(n\right)}{(n+\alpha)\, n}\times\nonumber \\
 &  & \hspace{-20pt}\hspace{-20pt}\hspace{-20pt}\times A_{0,0}\left(x,-n\right)\left(\delta_{r,0}\frac{A_{1,0}\left(x,-n\right)}{A_{0,0}\left(x,-n\right)}\alpha+n\,\bar{R}_{r}\left(x,\alpha\right)\right).\nonumber 
\end{eqnarray}
We can now solve (\ref{eq:Analyticity-cond-transseries}) for each
$m$, and this was done for $m=1,2$, in the same manner as for the
perturbative series in the previous Section. We found that 
\begin{equation}
\beta_{+}\left(1\right)=\frac{3}{4},\,\beta_{-}(1)=\frac{1}{4},\,\beta_{\pm}\left(2\right)=\pm\frac{1}{4},
\end{equation}
and also 
\begin{eqnarray}
\phi_{k}^{(m,+)}(n) & = & \sum_{\ell=0}^{k-1}Q_{k,\ell}^{(m,+)}\frac{1}{n^{\ell}}\phi_{1}^{(m,+)}(n),\nonumber \\
\phi_{k}^{(m,-)}(n) & = & \sum_{\ell=0}^{k}Q_{k,\ell}^{(m,-)}\frac{1}{n^{\ell}}\phi_{0}^{(m,-)}(n),
\end{eqnarray}
where 
\begin{eqnarray}
\phi_{1}^{(m,+)}(n) & = & -\mathcal{P}^{(m)}\phi_{0}^{(0,+)}(n),\nonumber \\
\phi_{0}^{(m,+)}(n) & = & 0,\\
\phi_{0}^{(m,-)}(n) & = & \mathcal{P}^{(m)}\phi_{0}^{(0,-)}\left(n-1\right).\nonumber 
\end{eqnarray}
The numerical coefficients $\mathcal{P}^{(m)}$ are also determined:
\begin{eqnarray}
\mathcal{P}^{(1)} & = & \frac{\overline{\sigma_{1}}\mathrm{e}^{3\pi\mathrm{i}/4}}{2}\frac{\Gamma\left(3/4\right)}{\Gamma(5/4)},\nonumber \\
\mathcal{P}^{(2)} & = & -\left(\mathcal{P}^{(1)}\right)^{2}.
\end{eqnarray}
It is worth noting that these coefficients seem to depend on the transseries
parameter, which it is not yet fixed. The value of the transseries
parameters can vary with the value of the coupling $x$: even if we
fix them to a particular value, when we move on the $x$-plane, these
parameters will jump in value when crossing any Stokes line (lines
where there are singularities on the Borel plane) -- such jump will
be governed by the so-called Stokes automorphism. So how can we interpret
the numerical values $\mathcal{P}^{(m)}$ if they depend on a parameter
which will change its value? In fact, the analyticity conditions
are solved in a specific direction on the $x$-plane: the one which
was chosen to perform the asymptotic expansions. In this region we
will have a specific value $\sigma_{1}\equiv\overline{\sigma_{1}}$,
and the $\mathcal{P}^{(m)}$ will be fixed at that position on the
$x$-plane.

Once the $c_{\pm}^{(m)}(n,x)$ are determined for $m=0,1,2$, we can
write down the corresponding transseries for the cusp anomalous dimension
(\ref{eq:Cusp-definition}). One can write down the full transseries
solution corresponding to (\ref{eq:Transseries-cpm}), but for the
present we will take $m_{2}=0$. Then
\begin{equation}
\Gamma_{\textrm{cusp}}\left(g,\sigma\right)=2g\left(1+\sum_{m=0}^{+\infty}\sigma^{m}\mathrm{e}^{-mA\frac{x}{2}}\Gamma^{(m)}(x)\right),\label{eq:cusp-transseries}
\end{equation}
where $\sigma\equiv\sigma_{1}\mathcal{P}^{(1)}$, $\Gamma^{(0)}$
was the perturbative contribution previously calculated (\ref{eq:Cusp-pert-sol}).
The two first non-perturbative corrections can be written as ($m=1,2$)
\begin{equation}
\Gamma^{(m)}(x)\simeq x^{-\beta\left(m\right)}\sum_{k=0}^{+\infty}\Gamma_{k}^{(m)}\left(\frac{x}{2}\right)^{-k},\label{eq:cusp-transseries-coeffs}
\end{equation}
where $\beta\left(m\right)=-\frac{m}{2}$ and the coefficients are%
\footnote{The first coefficients around the $m=1$ sector are in agreement with
\cite{Basso:2009gh}.%
}
\begin{eqnarray}
\Gamma_{0}^{(m)} & = & -4\frac{\mathcal{P}^{(m)}}{\left(\mathcal{P}^{(1)}\right)^{m}},
\end{eqnarray}
 
\begin{eqnarray}
\Gamma_{k}^{(m)} & = & -2^{1-k}\frac{\mathcal{P}^{(m)}}{\left(\mathcal{P}^{(1)}\right)^{m}}\times\\
 &  & \times\left(\sum_{a=0}^{k}\sum_{s=0}^{a}S_{-}\left(s\right)Q_{k-s,a-s}^{(m,-)}\sum_{r=0}^{a}4^{a+1-r}K_{r}^{(0-)}\right.+\nonumber \\
 &  & +\left.\sum_{a=0}^{k-1}\sum_{s=0}^{a}S_{+}\left(s\right)Q_{k-s,a-s}^{(m,+)}K_{a+1}^{(0+)}\right),\: k\ge1.\nonumber 
\end{eqnarray}

We have now calculated the perturbative coefficients around the first
two non-perturbative sectors. If we analyse the growth of these coefficients,
we find the same factorial growth $\Gamma\left(k-1/2\right)$ for
$m=0,1$, and the factorial growth $\Gamma\left(k+1/2\right)$ for
$m=2$: not only the perturbative series is asymptotic, but so are
the non-perturbative ones. Moreover the singularities which lead to
these sectors lie on the positive real axis, and thus one cannot properly
define a single integration contour on which to perform the resummation
of the Borel transform. 

Nevertheless, many lessons have been learned by now on the cases of
resurgent transseries: if our transseries is resurgent, then there
is a way of defining a single non-ambiguous result which properly cancels
the imaginary ambiguity at all non-perturbative orders. But in order
to use these results from resurgence, we first need to check that
our transseries is indeed resurgent.

\section{Resurgence and Large-Order}

Let us now check that the transseries formed by the asymptotic series
$\Gamma^{(m)}$ is indeed resurgent, \textit{i.e.} that the coefficients
$\Gamma_{k}^{(m)},\Gamma_{\ell}^{(m')}$ of neighbouring sectors $m,m'$
are related. To perform this check we use the coefficients $\Gamma_{k}^{(m)}$
of the asymptotic series (\ref{eq:cusp-transseries-coeffs}) and check
if their large order behaviour coincides with the large-order relations
predicted by resurgence techniques. These are relations between the
large order $k\gg1$ $\Gamma_{k}^{(m)}$ of one sector with the low
order coefficients of a nearby sector $\Gamma_{\ell}^{(m')}$. 

Take the transseries (\ref{eq:cusp-transseries}); assuming this transseries
is resurgent, we can use the so-called alien calculus to determine
the discontinuity of each asymptotic series $\Gamma^{(m)}$ across
singular directions (or Stokes lines). In our case (given we have
taken $m_{2}=0$) we only have one singular direction: the positive
real axis $\theta=0$. Resurgence then tells us \cite{Aniceto:2011nu}
that the discontinuity of the perturbative series along this direction
is (recall that $A=1/2$) 
\begin{equation}
\mathrm{Disc}_{0}\Gamma^{(0)}(x)=-S_{1}\mathrm{e}^{-A\frac{x}{2}}\Gamma^{(1)}-\left(S_{1}\right)^{2}\mathrm{e}^{-2A\frac{x}{2}}\Gamma^{(2)}+\cdots.
\end{equation}
There is one unknown constant in the above relation, the Stokes constant
$S_{1}$. As we will see, the large order relations will allow us
to determine this constant with great accuracy. This step is extremely
important as the Stokes constant plays a crucial role in the ambiguity
cancelation and resummation.

From the discontinuity, we can use Cauchy's theorem to determine large-order
relations \cite{ZinnJustin:1980uk}. Schematically, one writes
\begin{eqnarray}
\Gamma^{(0)}(z) & = & \oint_{\omega=z}\frac{d\omega}{2\pi\mathrm{i}}\frac{\Gamma^{(0)}(\omega)}{\omega-z}\\
 & \simeq & \int_{0}^{+\infty}\frac{d\omega}{2\pi\mathrm{i}}\frac{\mathrm{Disc}_{0}\Gamma^{(0)}(\omega)}{\omega-z}+\oint_{\infty}(\cdots).\nonumber 
\end{eqnarray}
In certain conditions, it can be shown by scaling arguments that the
integral at infinity does not contribute \cite{Bender:1990pd,Collins:1977dw}.
Expanding the r.h.s for large $z$, using the resurgence relation
for the discontinuity, and finally comparing equal powers of $z$
for the expansions in both sides of the equation, we arrive at the
relation
\begin{eqnarray}
\Gamma_{k}^{(0)} & \simeq & -\frac{S_{1}\mathcal{P}^{(1)}}{2\pi\mathrm{i}}\sum_{h=0}^{+\infty}\Gamma_{h}^{(1)}\frac{\Gamma\left(k-\frac{1}{2}-h\right)}{A^{k-\frac{1}{2}-h}}-\label{eq:Large-order-pert}\\
 &  & -\frac{\left(S_{1}\mathcal{P}^{(1)}\right)^{2}}{2\pi\mathrm{i}}\sum_{h=0}^{+\infty}\Gamma_{h}^{(1)}\frac{\Gamma\left(k-1-h\right)}{(2A)^{k-1-h}},\: k\gg1.\nonumber 
\end{eqnarray}
This formula states that if resurgence is expected, then the large
$k$ behaviour of the perturbative series is dictated by the coefficients
of the first non-perturbative sector, and then, more exponentially
suppressed ($2^{-k}$), the coefficients of the second non-perturbative
sector appear, and so on. The proportionality constant is once again
the Stokes constant $S_{1}$. Taking the ratio (which removes the
dependence on the yet unknown Stokes constant) of two consecutive
coefficients, and assuming $k\gg1$, we have a series (asymptotic
again)
\begin{equation}
\frac{\Gamma_{k}^{(0)}}{\Gamma_{k+1}^{(0)}}\frac{k}{A}\simeq\sum_{h=0}^{+\infty}c_{h}\, k^{-h},
\end{equation}
where the coefficients $c_{h}$ can be predicted from the original
large order relation (\ref{eq:Large-order-pert}). The first coefficients
are 
\begin{equation}
c_{0}=1;\: c_{1}=\frac{1}{2};\: c_{2}=\frac{1}{4}+A\frac{\Gamma_{1}^{(1)}}{\Gamma_{0}^{(1)}};\:\cdots.
\end{equation}
We can now check the convergence of $\Gamma_{k}^{(0)}$ to the coefficients
$c_{h}$ by successively removing the previous coefficient from the
ratio. For example to check the convergence to the coefficient $c_{2}$
we analyse 
\begin{equation}
\left(\left(\frac{\Gamma_{k}^{(0)}}{\Gamma_{k+1}^{(0)}}\frac{k}{A}-c_{0}\right)k-c_{1}\right)k\rightarrow c_{2}+O\left(k^{-1}\right).
\end{equation}

\begin{figure}
\begin{centering}
\includegraphics[width=1\columnwidth]{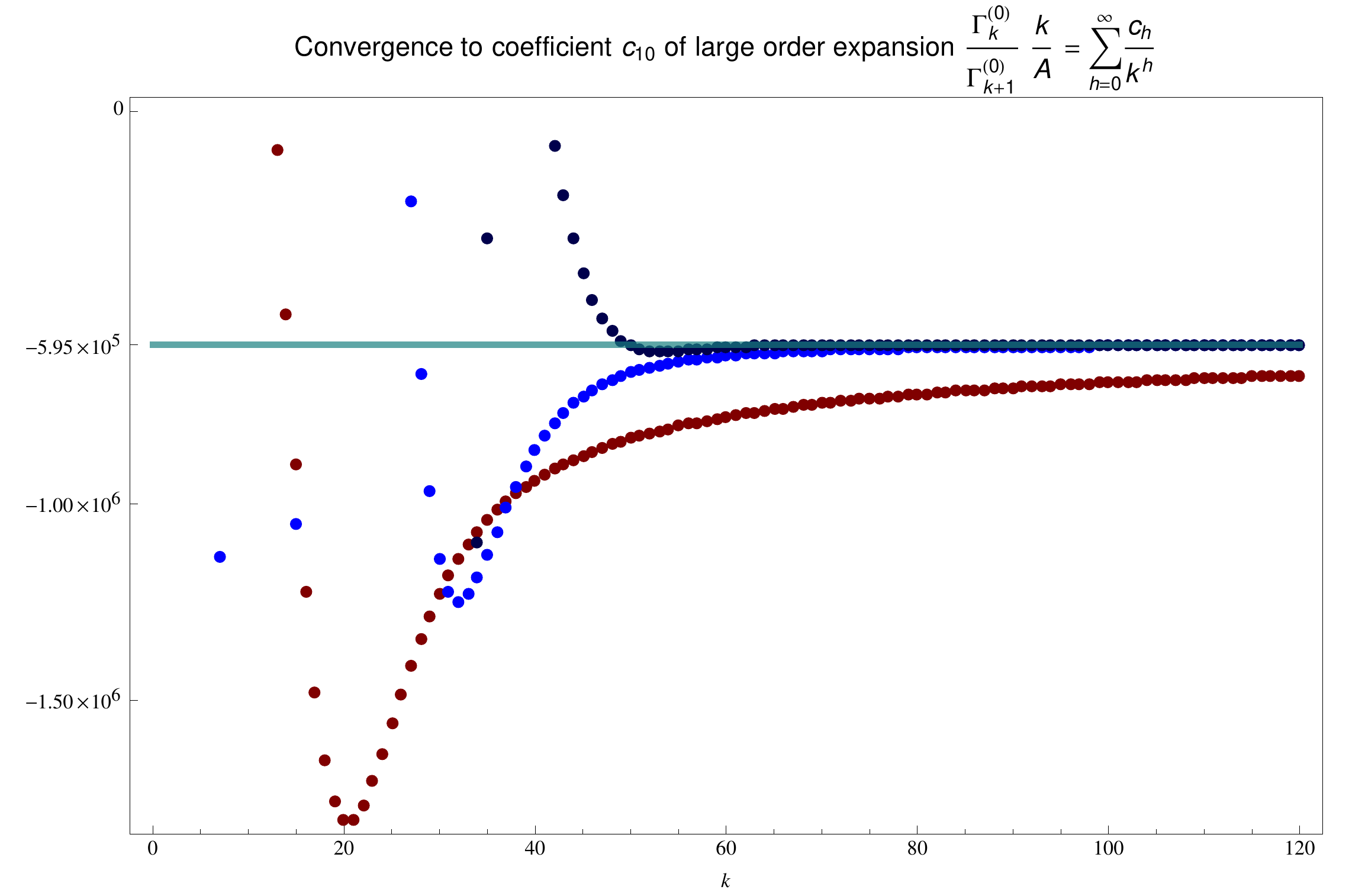}
\par\end{centering}

\protect\caption{Convergence of the large order ratio of perturbative coefficients
to the predicted result related to the first non-perturbative sector.
In red is shown the original ratio, and in blue the corresponding
Richardson transforms of order 2 (light blue) and 6 (dark blue). In
light green the predicted value for the coefficient $c_{10}$ is shown.\label{fig:Large-order-resurgence-check}}
\end{figure}

In Figure \ref{fig:Large-order-resurgence-check} we present the convergence
to coefficient $c_{10}$. In order for this convergence to be correct,
all of the previous coefficients need to be correct to a very high
accuracy, since factorial errors propagate rapidly. In this figure,
the original ratio (in red) is shown, together with two related Richardson
transforms which speed the convergence of this series in $1/n$ (see
\cite{Marino:2007te,Garoufalidis:2010ya}). The error between the
numerically calculated coefficient (via Richardson transforms) and
the predicted result from the large-order formula is of order $10^{-7}$. 

\begin{figure}
\begin{centering}
\includegraphics[width=1\columnwidth]{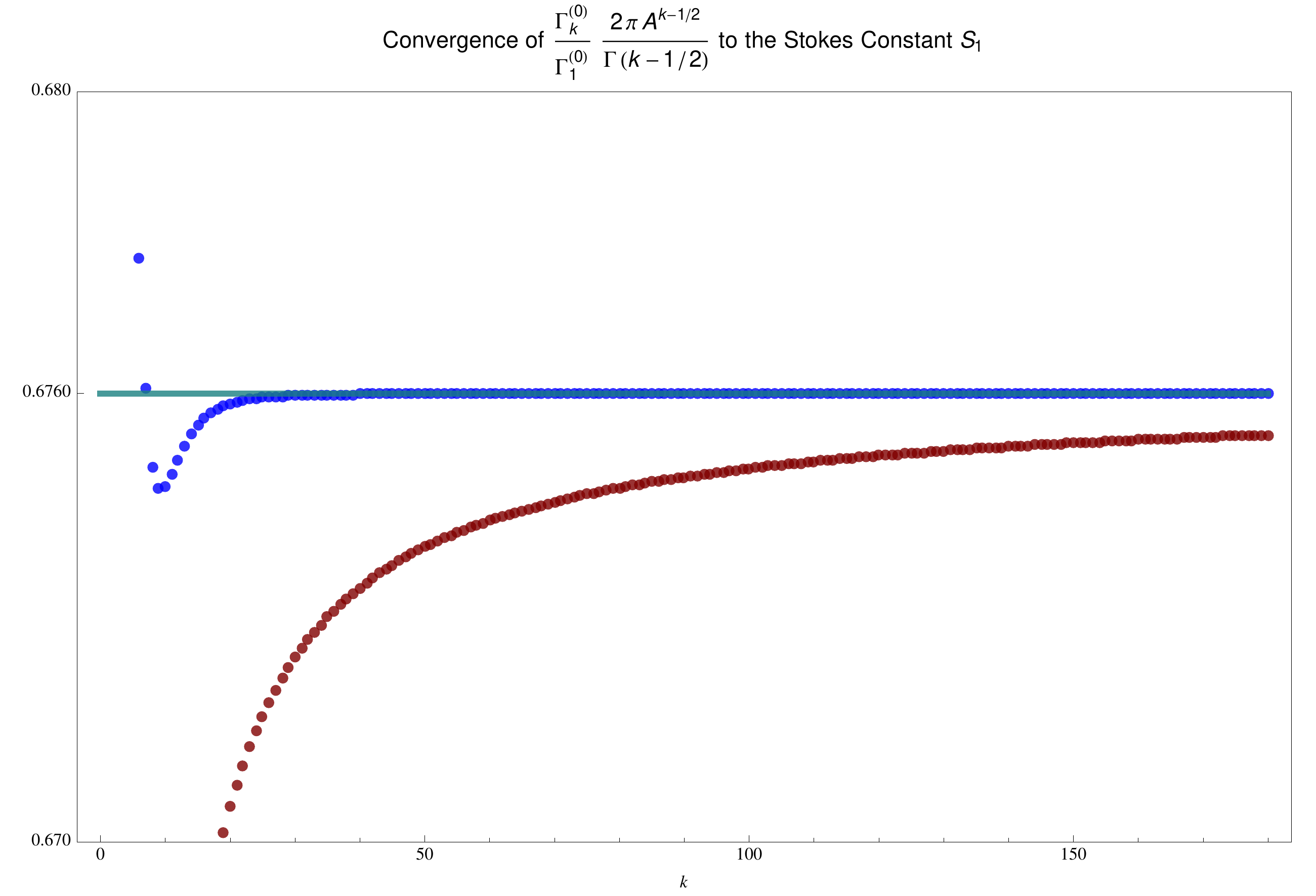}
\par\end{centering}

\protect\caption{Convergence of the perturbative coefficients to the Stokes constant.
In red the original series is shown, and in blue the second Richardson
transform with increased convergence. In dark green the calculated
value for the Stokes constant is shown.\label{fig:Large-order-Stokes-calculation}}
\end{figure}

If instead of dealing with the ratio of coefficients we analyse the
following 
\begin{eqnarray}
\frac{2\pi\, A^{k-\frac{1}{2}}\,\Gamma_{k}^{(0)}}{\Gamma\left(k-\frac{1}{2}\right)\Gamma_{0}^{(1)}} & \simeq & \mathrm{i}S_{1}\mathcal{P}^{(1)}\sum_{h=0}^{+\infty}\frac{\Gamma_{h}^{(1)}}{\Gamma_{0}^{(1)}}\frac{\Gamma\left(k-\frac{1}{2}-h\right)}{\Gamma\left(k-\frac{1}{2}\right)}A^{h}\nonumber \\
 & \sim & \mathrm{i}S_{1}\mathcal{P}^{(1)}+O\left(k^{-1}\right),
\end{eqnarray}
we directly obtain a convergence to the unknown Stokes constant. In
Figure \ref{fig:Large-order-Stokes-calculation} this convergence
is shown, with both the original series and a related Richardson transform.
The increase in convergence speed from the Richardson interpolation
method allows us to determine the Stokes constant to a very high accuracy.
Up to an error of $10^{-23}$ we find
\begin{equation}
\mathrm{i}S_{1}\mathcal{P}^{(1)}=\frac{1}{2}\frac{\Gamma\left(3/4\right)}{\Gamma(5/4)}.\label{eq:Stokes-constant-number}
\end{equation}

The same ideas were repeated for the asymptotic series of $\Gamma^{(1)}$,
whose large order will be directly related to coefficients of $\Gamma^{(2)}$,
finding that resurgence predictions also worked in this case. Note
that unlike the previous cases, the large-order behaviour of $\Gamma^{(2)}$
is dictated not only by $\Gamma^{(3)}$ but also by $\Gamma^{(1)}$
(the two nearest singularities on the Borel plane will be equally
distant from the origin, at $s=\pm A$). We conclude that indeed the
transseries for the cusp anomalous dimension is resurgent, and thus
we can apply the methods of ambiguity cancelation known to exist for
resurgent transseries. 

\section{Ambiguity Cancelation and Interpolation}

The next two questions which follow are: if our transseries is resurgent,
can we use this knowledge to write a non-ambiguous result? And if
this is possible, can we then interpolate from the strong coupling
asymptotic expansion to small coupling?

The answer to the first question is simply yes. The fact that we have
a resurgent transseries directly tells us how to obtain a non-ambiguous
result even when resumming in directions which are non-Borel summable,
\textit{i.e.}, along Stokes lines.

In order to verify the ambiguity cancelation it is sufficient to check
that the imaginary part of the resummed lateral transseries cancels
to higher and higher orders. The ambiguity coming from the perturbative
series, \textit{i.e.} the imaginary contribution from performing
a lateral resummation, is of order $\mathrm{e}^{-A\,4\pi g}$ for
each value of the coupling $g$. This is exactly the order at which
the first non-perturbative sector starts contributing. In fact, if
we add the two with the proper choice of parameter $\sigma$ in (\ref{eq:intro-transseries})
we can see that the newfound imaginary part will now be of order $\mathrm{e}^{-2A\,4\pi g}$
- the order of the second non-perturbative sector. In order to see this
cancelation, and to determine the value of the parameter $\sigma=\sigma_{1}\mathcal{P}^{(1)}$
which brings about the cancelation, we first need to resum our asymptotic
series. 

\begin{figure}
\begin{centering}
\includegraphics[width=1\columnwidth]{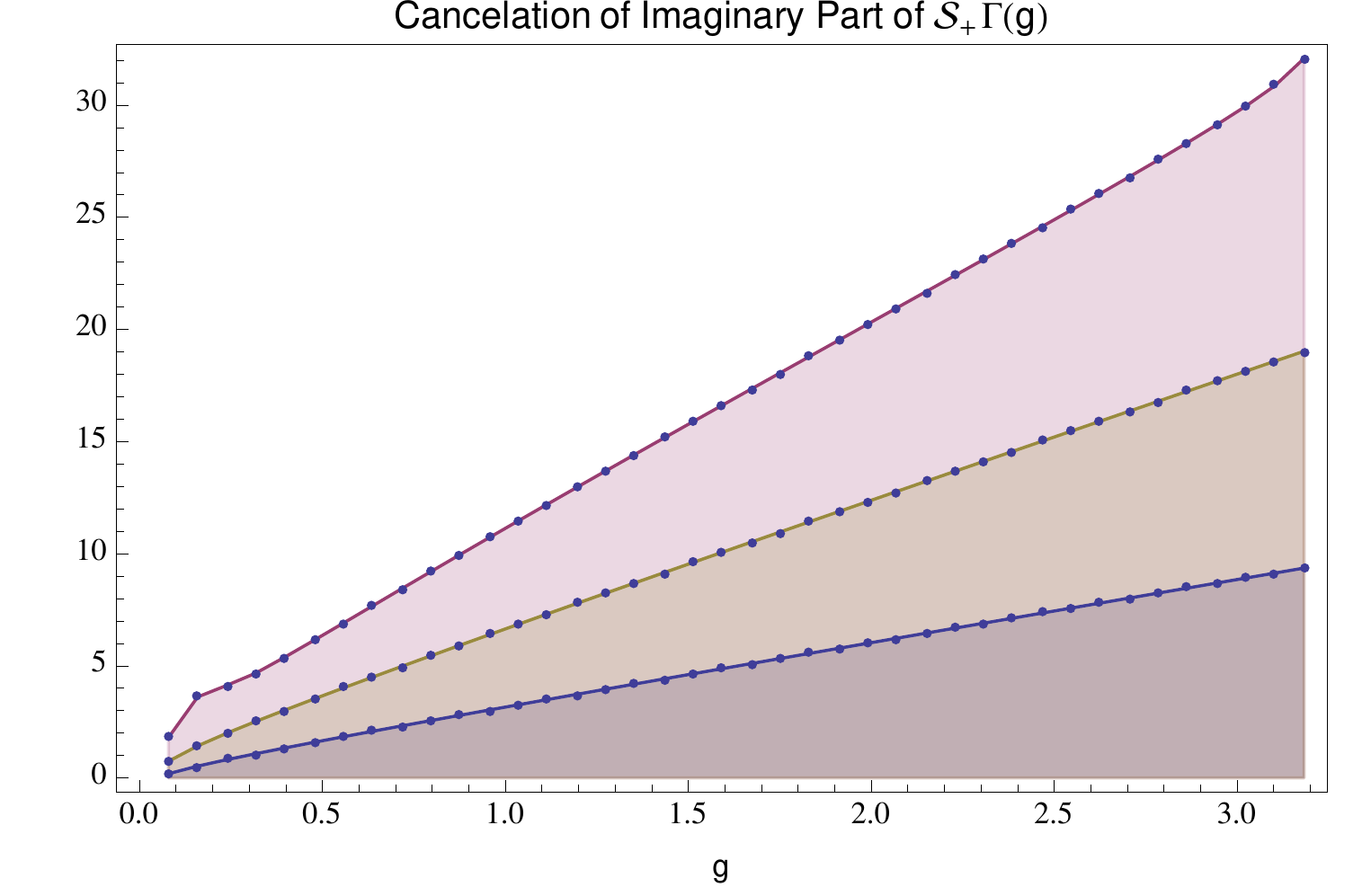}
\par\end{centering}

\protect\caption{Order of magnitude $10^{-\alpha}$ (where $\alpha$ is shown on the
y-axis) of the imaginary part of the transseries for the cusp anomalous
dimension, when one adds only the perturbative series (dark brown),
if one adds also the first non-perturbative correction (light brown)
and if one takes all three calculated contributions (light purple).\label{fig:Ambiguity-cancelation}}
\end{figure}

The method for resumming our series is the so-called \'{E}calle-Borel--Pad\'{e}
resummation method. It consists in calculating the Borel transforms
for each sector $\mathcal{B}\left[\Gamma^{(m)}\right](s)$, then determining
a Pad\'{e} approximant for each Borel transform, and subsequently
perform a lateral Borel resummation (we have chosen $\mathcal{S}_{+}$
as in (\ref{eq:Lateral-resums})) for different values of the coupling
$0.1\leq g\le4$ in order to obtain a resummed result for the full
transseries for general values of $g$.%
\footnote{It is very important to keep track of the first few terms of the series
which might not be included in the Borel transform, and thus need
to be added at a later stage. Also the overall factor can be left
out until after the resummation.%
} The imaginary part of the transseries for the cusp 
\begin{equation}
\frac{\mathcal{S}_{+}\Gamma}{2g}-1=\mathcal{S}_{+}\Gamma^{(0)}+\sigma\,\mathrm{e}^{-\frac{1}{2}4\pi g}\mathcal{S}_{+}\Gamma^{(1)}+\sigma^{2}\,\mathrm{e}^{-4\pi g}\mathcal{S}_{+}\Gamma^{(2)}+\cdots
\end{equation}
is given by (\ref{eq:Imaginary-part-transseries}). In Figure \ref{fig:Ambiguity-cancelation}
the order $10^{-\alpha}$ of the imaginary part can be seen if we
only include $\mathcal{S}_{+}\Gamma^{(0)}$ (dark brown), or if we
include both perturbative and first non-perturbative contributions
(light brown), and finally if we include all calculated contributions
(up to second order, in purple). The cancelation is shown for a range
of values of the coupling, ranging from weak to strong coupling. For
example for $g=1.5$ if we include only the perturbative part, the
imaginary ambiguity is of order $10^{-5}$. Including the first non-perturbative
sector cancels this imaginary part to $10^{-10}$. Including all three
sectors cancels the imaginary part up to 16 decimal places. The value
of the transseries parameter used was
 
\begin{equation}
\sigma_{0}=-\left(1+\mathrm{i}\right)\frac{\mathrm{i}S_{1}\mathcal{P}^{(1)}}{2}=\frac{1}{2\sqrt{2}}\mathrm{e^{-3\pi\mathrm{i}/4}}\frac{\Gamma\left(3/4\right)}{\Gamma\left(5/4\right)},\label{eq:transseries-param}
\end{equation}
with the Stokes constant given by (\ref{eq:Stokes-constant-number}).
This is in complete agreement with the proportionality constant appearing
in front of the non-perturbative corrections in \cite{Basso:2009gh}. 

\begin{figure}
\begin{centering}
\includegraphics[width=1\columnwidth]{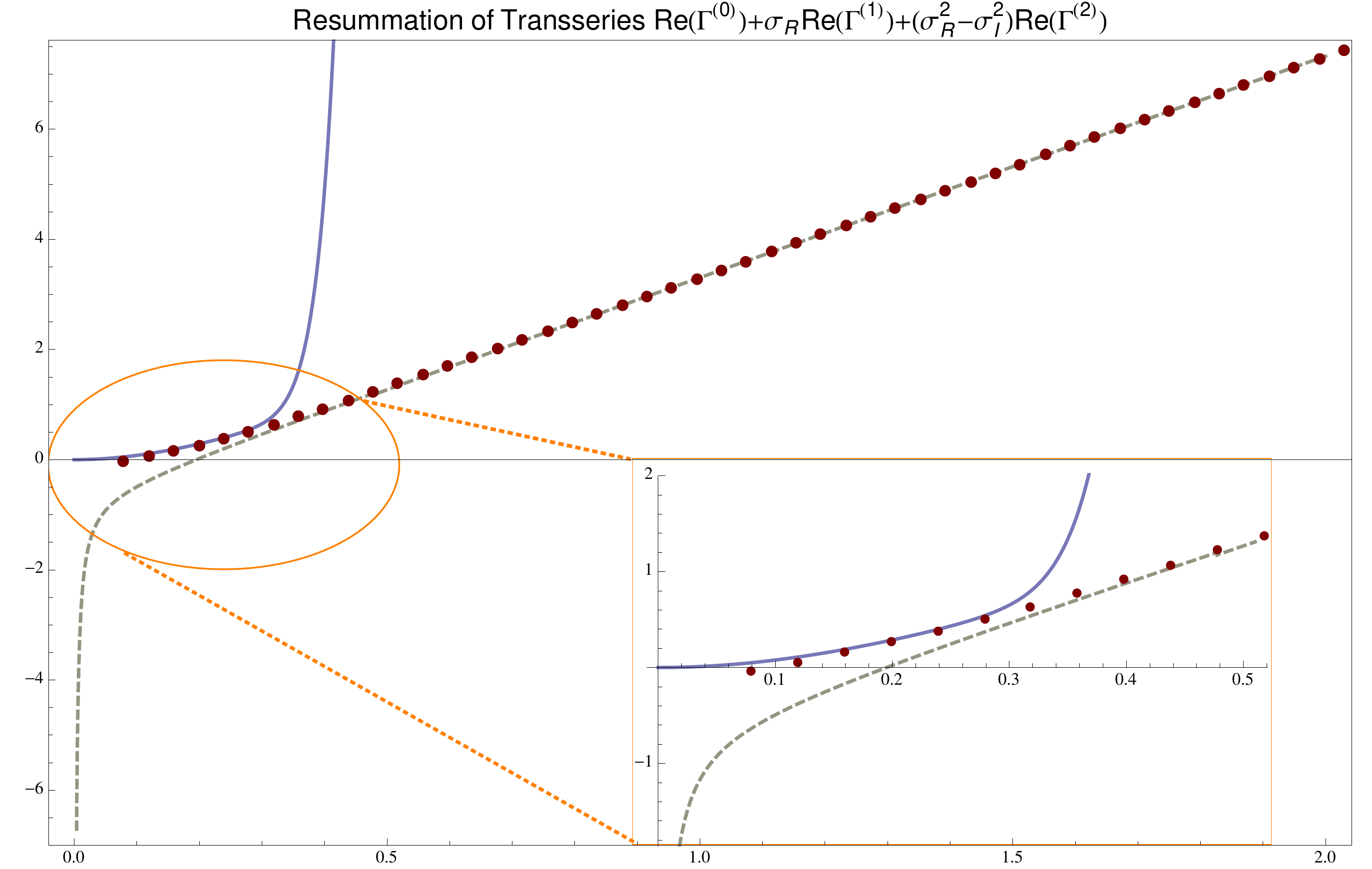}
\par\end{centering}

\protect\caption{Resummation of the transseries result for different values of the
coupling, shown in red. For large coupling the behaviour is dictated
by the perturbative series alone, as shown by the dashed line -- the
truncated summation of the perturbative expansion $\Gamma^{(0)}$,
which diverges for small g. In this regime another line is shown in
blue, corresponding to the small coupling expansion (up to 7 loops)
of the cusp anomalous dimension. The red dots interpolate between
these two regimes, starting to diverge around value $g=0.2$. \label{fig:Resummation}}
\end{figure}

We are now ready to answer the second question posed in this Section:
once we have managed to write down a real unambiguous transseries
result, given by (\ref{eq:Real-part-transseries}) with the parameter
$\sigma_{0}$ determined in (\ref{eq:transseries-param}), can we
use this result to interpolate between strong and weak coupling? In
Figure \ref{fig:Resummation} we show the truncated asymptotic series
result in a dashed green line: this result is accurate at strong coupling,
but diverges for weak coupling. In blue we plotted the weak coupling
expansion for the cusp anomalous dimension as determined (up to 7
loop order) by \cite{Beisert:2006ez}. Naturally this result diverges
for large values of the coupling. In red we show the resummed result
including the perturbative series and the first two non-perturbative
sectors. We see clearly that the red dots follow both the strong and
weak regime closely, starting to diverge for $g<0.2$. In order to
obtain more accurate results after this point, we would need to include
the next non-perturbative order.

In summary, once we have established resurgence of the transseries,
we are able to write down an unambiguous result, and use that same
result to reach values of the coupling as small as $g=0.2$. Moreover,
the information encoded in the transseries solution for the cusp anomalous
dimension goes beyond the strong/weak interpolation for positive real
values of the coupling. The resummation can be performed for $g\in\mathbb{C}$:
one can obtain a solution for any value of coupling, having in mind
that to reach certain values one might have to cross a Stokes line
and the transseries goes through a so-called Stokes transition (in
a Stokes transition the transseries parameters will jump in value,
and this jump is dictated by the Stokes constants and can be calculated
from resurgence techniques). In other words, the transseries solution
encodes the analytic properties of the cusp anomalous dimension as
a function of the coupling. 

Another extremely insightful example of how the transseries encoded
the analytic properties of the observable was recently achieved in
the context of matrix models \cite{Couso-Santamaria:2015wga}, where
from a large $N$ asymptotic expansion, resurgence and resummation
techniques allowed the authors to reach not only finite values of
the rank $N$, but they were also able to analytically continue their
results to any complex value of $N$, once Stokes phenomena were taken
into account.

\section{Conclusions/Outlook}

In this work we presented a thorough analysis of the resurgent properties
of the cusp anomalous dimension's strong coupling expansion. When
analysing the perturbative series we found that there are two types
of singularities on the Borel plane in both the positive and negative
real axis. These need to be taken into account in
order to fully solve the analyticity conditions coming from the BES
equations. Nevertheless their physical interpretation has not yet
been addressed. The non-perturbative behaviour associated with singularities
at $s=nA$ (on the positive real axis), were seen to be directly linked
to the mass gap of the $O(6)$ $\sigma$-model at least for $n=1$ \cite{Basso:2009gh}.
The singularities at $s=-4nA$ are much more suppressed and as such
haven't been addressed, but some questions can be immediately raised:
does the same feature appear for the energy density of the $O(6)$
model? Does the relation between scaling function and energy density
hold for higher exponentially suppressed contributions? Since the
two types of non-perturbative phenomena are collinear, will we witness
resonance?

For the aims of the current paper we used a transseries ansatz for
the cusp anomalous dimension which did not include the second type
of non-perturbative phenomena, as we only studied the solution up
to $n=2$ in the non-perturbative order. Nevertheless this was enough
to check the resurgent properties of the strong coupling expansion,
with the large-order relations predicted by resurgence accurately
solving the large order behaviour of the perturbative expansion and
first non-perturbative order.

In our case we could determine the asymptotic expansions around the
non-perturbative sectors via the BES equation, and use these
to check the resurgence of the transseries. In cases when only perturbation
theory is known, one can go the opposite way and use the predictions
of resurgence to determine the coefficients of the expansions around
the non-perturbative sectors.

Knowing that the transeries proposed is indeed resurgent we then proceeded
to determine the resummed transseries, using a lateral resummation
procedure. This naturally introduces an imaginary ambiguity, which
can then be seen to cancel given the proper choice of the transseries
parameter: using the methods of median resummation \cite{Aniceto:2013fka}
we determined the transseries parameter to be the one proposed in
\cite{Basso:2009gh}.

Finally the resummation procedure can be done for different values
of the coupling, and we showed that including up to second order non-perturbative
effects, we could systematically obtain accurate results for the cusp
anomalous dimension all the way from strong coupling up to $g\simeq0.2$.
Moreover we can perform the resummation for any values of the coupling,
as long as we take into consideration the Stokes phenomena occurring
when we analytically continue our results across Stokes lines.

This work is an example in the use and elegance of the resurgence techniques.
Knowing the perturbative asymptotic expansion of an observable
in some regime, we can determine the non-perturbative corrections
via large order relations, and upgrade the solution to a transseries.
We can then use resummation methods such as \'{E}calle-Borel--Pad\'{e}
to obtain a resummed solution for any value of the coupling (using
resurgence techniques to analytically continue the solution across
Stokes lines). This resummed transseries encodes the analytic properties
of our observable as a function of the coupling.

Other open questions still to be addressed are whether one can use
a transseries ansatz for the auxiliary function in the BES equation,
and through this find an equivalent approach to determining the coefficients $\Gamma_{k}^{(m)}$
to the one used in \cite{Basso:2007wd,Basso:2009gh}. Also, it remains
to be understood how these results translate to the problem of the
scaling function $\epsilon(g,j)$ and the energy density of the $O(6)$
$\sigma$-model. The first question is if one can find a full transseries
in these cases. If so, will we have the same type of singularities?
Finally, in these two cases we have two parameters $g,j$ and it would
be important to understand how the different regimes dictated by these
parameters appear in the resurgent context.
\begin{acknowledgments}
I would like to thank Ricardo Schiappa, Romuald Janik and Hesam Soltanpanahi
for useful comments and reading of an earlier version of this manuscript.
This work was supported by the NCN grant 2012/06/A/ST2/00396.
\end{acknowledgments}

\appendix

\section{Asymptotic Expansions\label{sec:App-Asymptotic-Expansions}}

In order to write down the asymptotic expressions appearing in the
main text, we need first to define the following four asymptotic expansions

\begin{eqnarray}
A_{0,0}\left(x,\alpha\right) & = & \sum_{s=0}^{+\infty}\frac{\Gamma\left(\frac{3}{4}+s\right)\Gamma\left(-\frac{1}{4}+s\right)}{\Gamma\left(-\frac{1}{4}\right)\, s!\,\left(x\,\alpha\right)^{s}},\\
A_{0,1}\left(x,\alpha\right) & = & \sum_{s=0}^{+\infty}\frac{\Gamma\left(\frac{1}{4}+s\right)\Gamma\left(\frac{5}{4}+s\right)}{\Gamma\left(\frac{1}{4}\right)\, s!\,\left(x\,\alpha\right)^{s}},
\end{eqnarray}
\begin{eqnarray}
A_{1,0}\left(x,\alpha\right) & = & \sum_{s=0}^{+\infty}\frac{\Gamma\left(\frac{3}{4}+s\right)^{2}}{\Gamma\left(\frac{3}{4}\right)\, s!\,\left(x\,\alpha\right)^{s}},\\
A_{1,1}\left(x,\alpha\right) & = & \sum_{s=0}^{+\infty}\frac{\Gamma\left(\frac{1}{4}+s\right)^{2}}{\Gamma\left(\frac{1}{4}\right)\, s!\,\left(x\,\alpha\right)^{s}}.
\end{eqnarray}
In all the expressions in this appendix we assume $\alpha\ne0$. The
functions $U_{n}^{\pm}\left(x\right),\, n=0,1$ have been defined
in \cite{Basso:2009gh} in terms of Whittaker functions of the second
kind. For the purposes of this paper however, we are only interested
in their asymptotic expansions for large $x$, which are:
\begin{eqnarray}
U_{0}^{+}\left(\frac{nx}{2}\right) & \simeq & \left(nx\right)^{-5/4}A_{0,1}\left(x,-n\right),\nonumber \\
U_{0}^{-}\left(\frac{nx}{2}\right) & \simeq & \left(nx\right)^{-3/4}A_{0,0}\left(x,-n\right),\\
U_{1}^{+}\left(\frac{nx}{2}\right) & \simeq & \frac{1}{2}\left(nx\right)^{-1/4}A_{1,1}\left(x,-n\right),\nonumber \\
U_{1}^{-}\left(\frac{nx}{2}\right) & \simeq & \frac{1}{2}\left(nx\right)^{-3/4}A_{1,0}\left(x,-n\right).\nonumber 
\end{eqnarray}
Note that even from these asymptotic expansions it is not difficult
to see that these functions have a cut on the negative real axis.
The other functions of interest appearing in the main text are $V_{n}\left(x\right),\, n=0,1$.
Again, in \cite{Basso:2009gh} these are entire functions which
were written in terms of Whittaker functions of the first kind, but
for our purposes we are only interested in their asymptotic expansions.
More specifically we will only need the asymptotic expansion of their
ratio:
\begin{equation}
2r\left(\alpha\right)\equiv2\frac{V_{1}\left(\frac{\alpha x}{2}\right)}{V_{0}\left(\frac{\alpha x}{2}\right)}.
\end{equation}

The asymptotic expansion of this ratio for large $x$ depends on the
sign of $\alpha$. For $\alpha=\left|\alpha\right|>0$
\begin{equation}
2r\left(\left|\alpha\right|\right)\simeq\sum_{k=0}^{+\infty}(-1)^{k}\mathrm{e}^{-k\left|\alpha\right|x}\mathrm{e}^{5\pi\mathrm{i}k/4}\left(\left|\alpha\right|x\right)^{-k/2}R_{k}\left(x,\left|\alpha\right|\right)
\end{equation}
where 
\begin{eqnarray}
R_{0}\left(x,\left|\alpha\right|\right) & = & \frac{A_{1,0}\left(x,\left|\alpha\right|\right)}{A_{0,0}\left(x,\left|\alpha\right|\right)},\\
R_{k}\left(x,\left|\alpha\right|\right) & = & \left(R_{0}\left(x,\left|\alpha\right|\right)+\left|\alpha\right|x\frac{A_{1,1}\left(x,-\left|\alpha\right|\right)}{A_{0,1}\left(x,-\left|\alpha\right|\right)}\right)\times\nonumber \\
 &  & \qquad\qquad\qquad\qquad\times\left(\frac{A_{0,1}\left(x,-\left|\alpha\right|\right)}{A_{0,0}\left(x,\left|\alpha\right|\right)}\right)^{k}.\nonumber 
\end{eqnarray}
For $\alpha=-\left|\alpha\right|<0$ we then have 
\begin{equation}
2r\left(-\left|\alpha\right|\right)\simeq\sum_{k=0}^{+\infty}(-1)^{k}\mathrm{e}^{-k\left|\alpha\right|x}\mathrm{e}^{3\pi\mathrm{i}k/4}\left(\left|\alpha\right|x\right)^{k/2}\widetilde{R}_{k}\left(x,\left|\alpha\right|\right)
\end{equation}
where 
\begin{eqnarray}
\widetilde{R}_{0}\left(x,\left|\alpha\right|\right) & = & x\left|\alpha\right|\frac{A_{1,1}\left(x,\left|\alpha\right|\right)}{A_{0,1}\left(x,\left|\alpha\right|\right)},\\
\widetilde{R}_{k}\left(x,\left|\alpha\right|\right) & = & \left(\widetilde{R}_{0}\left(x,\left|\alpha\right|\right)-\frac{A_{1,0}\left(x,-\left|\alpha\right|\right)}{A_{0,0}\left(x,-\left|\alpha\right|\right)}\right)\times\nonumber \\
 &  & \qquad\qquad\qquad\qquad\times\left(\frac{A_{0,0}\left(x,-\left|\alpha\right|\right)}{A_{0,1}\left(x,\left|\alpha\right|\right)}\right)^{k}.\nonumber 
\end{eqnarray}

In Section \ref{sec:Transseries-and-Analyticity} when solving the
analyticity conditions, some particular combinations of the asymptotic
expansions $A_{i,j}\left(x,\alpha\right)$ repeatedly appeared. These
were
\begin{eqnarray}
M_{0}^{(+)}(x) & \equiv & \frac{A_{0,1}\left(x,-n\right)}{x}\times\nonumber \\
 &  & \qquad\times\left(\frac{A_{1,1}(x,-n)}{A_{0,1}\left(x,-n\right)}\left|\alpha\right|x+\frac{A_{1,0}\left(x,\left|\alpha\right|\right)}{A_{0,0}\left(x,\left|\alpha\right|\right)}\right)\nonumber \\
 & \simeq & \left|\alpha\right|\Gamma\left(\frac{5}{4}\right)\sum_{r=0}^{+\infty}x^{-r}\sum_{\ell=0}^{r}\frac{M_{r,\ell}^{(0,+)}}{n^{\ell}\left|\alpha\right|^{r-\ell}},
\end{eqnarray}
\begin{eqnarray}
M_{1}^{(+)}(x) & \equiv & \frac{A_{0,0}\left(x,-n\right)}{xn\left(n+\left|\alpha\right|\right)}\times\nonumber \\
 &  & \qquad\times\left(\frac{A_{1,0}(x,-n)}{A_{0,0}\left(x,-n\right)}\left|\alpha\right|+n\frac{A_{1,0}\left(x,\left|\alpha\right|\right)}{A_{0,0}\left(x,\left|\alpha\right|\right)}\right)\nonumber \\
 & \simeq & \frac{1}{n}\Gamma\left(\frac{3}{4}\right)\sum_{r=0}^{+\infty}x^{-r-1}\sum_{\ell=0}^{r}\frac{M_{r,\ell}^{(1,+)}}{n^{\ell}\left|\alpha\right|^{r-\ell}},
\end{eqnarray}
\begin{eqnarray}
M_{0}^{(-)}(x) & \equiv & \frac{A_{0,1}\left(x,-n\right)}{(n+\left|\alpha\right|)}\left|\alpha\right|\times\nonumber \\
 &  & \qquad\times\left(-\frac{A_{1,1}(x,-n)}{A_{0,1}\left(x,-n\right)}+\frac{A_{1,1}\left(x,\left|\alpha\right|\right)}{A_{0,1}\left(x,\left|\alpha\right|\right)}\right)\nonumber \\
 & \simeq & \frac{1}{n}\Gamma\left(\frac{5}{4}\right)\sum_{r=0}^{+\infty}x^{-r-1}\sum_{\ell=0}^{r}\frac{M_{r,\ell}^{(0,-)}}{n^{\ell}\left|\alpha\right|^{r-\ell}},
\end{eqnarray}
\begin{eqnarray}
M_{1}^{(-)}(x) & \equiv & \frac{A_{0,0}\left(x,-n\right)}{xn}\left|\alpha\right|\times\nonumber \\
 &  & \qquad\times\left(-\frac{A_{1,0}(x,-n)}{A_{0,0}\left(x,-n\right)}+xn\frac{A_{1,1}\left(x,\left|\alpha\right|\right)}{A_{0,1}\left(x,\left|\alpha\right|\right)}\right)\nonumber \\
 & \simeq & \left|\alpha\right|\Gamma\left(\frac{3}{4}\right)\sum_{r=0}^{+\infty}x^{-r}\sum_{\ell=0}^{r}\frac{M_{r,\ell}^{(1,-)}}{n^{\ell}\left|\alpha\right|^{r-\ell}}.
\end{eqnarray}

\section{Relations Between Sums\label{sec:App-Relations-Between-Sums}}

Given the two fundamental objects $\phi_{0}^{(0,\pm)}\left(n\right)$
independent of the coupling $x=8\pi g$, appearing in the transseries
of the coefficients $c_{\pm}\left(n,x\right)$ in (\ref{eq:Transseries-cpm}),
these can easily be determined from analyticity conditions to be 
\begin{eqnarray}
\phi_{0}^{(0,+)}(n) & = & 2\frac{\Gamma\left(n-\frac{3}{4}+1\right)}{n!\,\Gamma\left(\frac{1}{4}\right)^{2}},\\
\phi_{0}^{(0,-)}\left(n\right) & = & \frac{\Gamma\left(n-\frac{1}{4}+1\right)}{2\, n!\,\Gamma\left(\frac{3}{4}\right)^{2}}.
\end{eqnarray}
With these definitions we calcuate the following sums for $m\ge1$
\begin{eqnarray}
K_{m}^{(0,+)} & \equiv & -\Gamma\left(\frac{5}{4}\right)\sum_{n\ge1}\frac{\phi_{0}^{(0,+)}(n)}{n^{m}}\nonumber \\
 & = & -\frac{1}{8}\,_{m+2}F_{m+1}\left(\mathbf{a}^{+},\mathbf{b};1\right),\\
K_{m}^{(0,-)} & \equiv & -\Gamma\left(\frac{3}{4}\right)\sum_{n\ge1}\frac{\phi_{0}^{(0,-)}(n)}{n^{m}}\nonumber \\
 & = & -\frac{3}{8}\,_{m+2}F_{m+1}\left(\mathbf{a}^{-},\mathbf{b};1\right),
\end{eqnarray}
where $\,_{p+1}F_{p}$ is a generalized hypergeometric function, $\mathbf{a}^{\pm}$
are vectors with $m+2$ entries: $\mathbf{a^{+}}=\left(1,\cdots,1,5/4\right)$
and $\mathbf{a^{-}}=\left(1,\cdots,1,7/4\right)$, whereas $\mathbf{b}=\left(2,\cdots,2\right)$
is a vector with $m+1$ entries. Defining also $K_{0}^{(0,\pm)}\equiv1/2$,
one can easily see that 
\begin{eqnarray}
\Gamma\left(\frac{5}{4}\right)\sum_{n\ge1}\frac{\phi_{0}^{(0,+)}(n)\left|\alpha\right|}{(n-\left|\alpha\right|)n^{m}} & = & \sum_{\ell=0}^{m}\frac{K_{\ell}^{(0,+)}}{\left|\alpha\right|^{m-\ell}},\nonumber \\
\Gamma\left(\frac{3}{4}\right)\sum_{n\ge1}\frac{\phi_{0}^{(0,-)}(n)\left|\alpha\right|}{(n-\left|\alpha\right|)n^{m}} & = & \sum_{\ell=0}^{m}\frac{K_{\ell}^{(0,-)}}{\left|\alpha\right|^{m-\ell}}.
\end{eqnarray}
In all the expressions in this Appendix we assume $\alpha_{\ell}=\ell-1/4$,
for $\ell\in\mathbb{Z}$. Other useful identities of $\phi_{0}^{(0,-)}(n)$
are
\begin{equation}
\sum_{n\ge1}\frac{\phi_{0}^{(0,-)}(n-1)}{n-\left|\alpha_{\ell}\right|}=0,\:\ell\le-1,
\end{equation}
\begin{eqnarray}
\Gamma\left(\frac{3}{4}\right)\sum_{n\ge1}\frac{\phi_{0}^{(0,-)}(n-1)}{\left(n-\left|\alpha_{\ell}\right|\right)n^{m}} & =\\
 &  & \hspace{-105pt}=-\sum_{s=0}^{m-1}\frac{1}{\left|\alpha_{\ell}\right|^{m-s}}\sum_{r=0}^{s}\frac{K_{r}^{(0,-)}}{\left|\alpha_{0}\right|^{s-r+1}},\: m,-\ell\ge1,\nonumber 
\end{eqnarray}

\begin{equation}
\Gamma\left(\frac{3}{4}\right)\sum_{n\ge1}\frac{\phi_{0}^{(0,-)}(n-1)}{n^{m}}=\sum_{\ell=0}^{m-1}\frac{K_{\ell}^{(0,-)}}{\left|\alpha_{0}\right|^{m-\ell}},\: m\ge1,
\end{equation}
\begin{equation}
\Gamma\left(\frac{3}{4}\right)\left|\alpha_{0}\right|\sum_{n\ge1}\frac{\phi_{0}^{(0,-)}(n-1)}{(n-\left|\alpha_{0}\right|)n^{m}}=\sum_{\ell=0}^{m}\frac{K_{\ell}^{(1,-)}}{\left|\alpha_{0}\right|^{m-\ell}},\: m\ge0,
\end{equation}
where $K_{0}^{(1,-)}=\pi/(4\sqrt{2})$ and for $\ell\ge1$ $K_{\ell}^{(1,-)}=(-1)\sum_{a=0}^{\ell-1}4^{\ell-a}K_{a}^{(0,-)}$.
Useful identities of $\phi_{0}^{(0,+)}(n)$ are 
\begin{equation}
\Gamma\left(\frac{5}{4}\right)\left|\alpha_{0}\right|\sum_{n\ge1}\frac{\phi_{0}^{(0,+)}(n)}{(n+\left|\alpha_{0}\right|)n^{m}}=-\sum_{\ell=0}^{m}\left(-4\right)^{m-\ell}K_{\ell}^{(1,+)},
\end{equation}
where $K_{0}^{(1,+)}=\frac{1}{2}-\frac{\pi}{4\sqrt{2}}$ and $K_{\ell}^{(1,+)}=K_{\ell}^{(0,+)}$
for $\ell\ge1$.

\bibliographystyle{my-utphys}
\bibliography{complete-library-processed,resurgence_bib}

\end{document}